\documentclass[aps,pre,twocolumn,showpacs]{revtex4-1}
\usepackage{graphicx}
\usepackage{amsmath,amssymb,latexsym}
\usepackage{amsfonts}

\usepackage{epsfig}

\bibstyle{apsrev.bib}

\begin{document}

\newcommand{\mean}[1]{{\left< #1 \right>}}
\newcommand{\UN}{{\emptyset}}

\title{Universal properties of knotted polymer rings}
 
\author{M. Baiesi}
\author{E. Orlandini}
\affiliation{Dipartimento di Fisica e Astronomia, Universit\`a di Padova, Via Marzolo 8, I-35131 Padova, Italy}
\affiliation{Sezione INFN, Universit\`a di Padova, Via Marzolo 8, I-35131 Padova, Italy}

\date{\today}

\begin{abstract}
By performing Monte Carlo sampling of $N$-steps self-avoiding polygons embedded 
on different Bravais lattices we explore the robustness of universality in the entropic, 
metric and geometrical  properties of knotted polymer rings. 
In particular, by simulating polygons with $N$ up to $10^5$ we furnish a 
sharp estimate of the asymptotic values of the knot probability ratios and show 
their  independence on the lattice type. This universal feature was previously 
suggested although with different estimates of the asymptotic values.
In addition we show that the scaling behavior of the mean squared radius of gyration of 
polygons depends on their  knot type only through its correction to scaling. 
Finally, as a measure  of the geometrical self-entanglement of the SAPs 
we consider the standard deviation of the writhe distribution and estimate its power-law behavior 
in the large  $N$ limit. The estimates of the power exponent do depend neither on the lattice 
nor on the knot type, strongly supporting an extension of the universality  property to some features of the geometrical entanglement. 
\end{abstract}


\maketitle

\section{Introduction}
Scaling hypothesis and the renormalization group theory led to 
the understanding of universality in critical phenomena, i.e.~the fact that systems  
that look different at small scales may share, in proximity of critical points, 
common statistical properties. This concept greatly reduces the variety of possible 
critical behavior by grouping all systems into a small number of universality 
classes characterized by the same critical exponents and critical amplitude ratios. 
In this respect universality justifies the study of systems in a given universality class 
that are simple enough to be treated either numerically or analytically.

In statistics of polymer conformations this approach has been often followed by 
modeling polymers as self-avoiding walks (SAWs) on discrete lattices such as the simple cubic (SC), the face-centered-cubic (FCC) and the body-centered cubic lattice (BCC).  
For these models combinatorial arguments and 
Monte Carlo simulations can be efficiently applied to obtain rigorous and numerical  
results on  the large scale (asymptotic) behavior of polymers in good solution~\cite{Madras&Slade:1993}.
Moreover SAWs can be mapped to a magnetic system at its critical point and 
studied by renormalization group techniques~\cite{DeGennes:1979,Vanderzande:1998}. 
This approach has led, for example, to the well established results that the number $Z(N)$ and 
the mean squared radius of gyration $\langle R_g^2(N) \rangle_p$
of $N$-steps self-avoiding polygons 
(SAPs, i.e.~$(N-1)$-steps SAWs having the two extremities one lattice distance apart) 
behave for large $N$ respectively  as 
\begin{equation}
Z(N) \simeq A \mu^{N} N^{\alpha-2}\left (1+ \frac{a}{N^{\Delta}} + \cdots \right )
\label{scal_entropy_SAP}
\end{equation}
and
\begin{equation}
\langle R_g^2(N) \rangle_p \simeq B N^{2\nu} \left (1+ \frac{b}{N^{\Delta}} + \cdots \right ),
\label{scal_radius_SAP}
\end{equation}
where the amplitudes $A,a$, $B,b,$  and the connective constant $\mu$ are non universal quantities that 
depend on the underlying lattice~\cite{Madras&Slade:1993}, while 
$\alpha$ and $\nu$ are universal exponents related by the hyperscaling equation $\alpha = 2-\nu d$ with $d$ the dimensionality of the space and $\nu$ the metric exponent. 
In $d=3$, advanced numerical simulations give $\nu \simeq 0.587597(7)$~\cite{Clisby:2010:Phys-Rev-Lett}, and hence $\alpha \simeq 0.237209(21)$, in good agreement with field theoretical results~\cite{Guida&Zinn-Justin:1997:JPA}. 
Also the correction to scaling exponent $\Delta$ is believed
to be universal and  its best estimate is $\Delta = 0.528(12)$~\cite{Clisby:2010:Phys-Rev-Lett}.

Notice that SAPs on regular lattices  model 
polymer rings in good solution with \emph{any} possible  topology, i.e.~with an arbitrary 
number of knots tied in.  However, in most experiments and physical situations such as in 
melts and concentrated solutions,  
the overall topology of a system of ring polymers cannot be changed unless the excluded volume interaction is violated.
This, for example, has relevant consequences on the temperature of the collapse transition of ring polymers,
which is found experimentally to be different from that of their linear counterparts~\cite{Takano09}.

Because of the topological constraint for looped polymers, the above mentioned mapping to a 
magnetic system is not valid anymore and, consequently, no field theory argument can establish 
the validity of scaling laws similar to (\ref{scal_entropy_SAP}) and (\ref{scal_radius_SAP}).  
On the basis of previous numerical 
investigations~\cite{Orlandini:1998:J-Phys-A,Marcone:2007:PRE,Rawdon:Macromol:2008a,Rensburg&Rechnitzer:2008:JPA}
it is however reasonable to assume
\begin{equation}
Z_k(N) \simeq A_{k} \mu_k^{N} N^{\alpha_k-2}\left (1+ \frac{a_k}{N^{\Delta_k}} + \cdots \right )
\label{scal_entropy_knot}
\end{equation}
and
\begin{equation}
\langle R_g^2(N) \rangle_k \simeq B_{k} N^{2\nu_k} \left (1+ \frac{b_k}{N^{\Delta_k}} + \cdots \right ),
\label{scal_radius_knot}
\end{equation}
where $\mu_k$ and $\alpha_k$ are respectively the connective constant and the entropic exponent of the subset of SAPs with a given knot type $k$.
With this notation $k$ refers either to prime knots or to composite knots given by the connect 
sum of prime knots 
(roughly speaking, a knot is composite if it can be split in two knots located in different portions of the chain, which are separated by a plane)~\cite{Orlandini:1996:J-Phys-A,Orlandini&Whittington:2007:Rev-Mod_Phys}.  The case $k= \emptyset$ denotes the special case of unknotted
SAPs (SAPs with trivial topology).

The scaling relations (\ref{scal_entropy_knot}) and (\ref{scal_radius_knot}) have been conjectured 
in analogy with (\ref{scal_entropy_SAP}) and (\ref{scal_radius_SAP})  and their validity 
has been confirmed so far only by numerical simulations~\cite{Orlandini:1996:J-Phys-A,Orlandini:1998:J-Phys-A} 
with the findings  $\mu_k = \mu_{\emptyset}<\mu$,  $\alpha_k = \alpha_{\emptyset} + m_k$, 
$\nu_k=\nu$ and $B_k=B_{\emptyset}$ where $m_k$  denote the number of prime components in the knot 
decomposition~\cite{Orlandini:1996:J-Phys-A,Orlandini&Whittington:2007:Rev-Mod_Phys}.
The {\em probability} of occurrence of a given knot $k$,
\begin{equation}
P_k(N) = \frac{Z_k(N)}{Z(N)},
\label{prob_k}
\end{equation}
is thus dominated by the exponential decay $(\mu_{\emptyset}/\mu)^N$ at large $N$s. 
In the following we will not deal with this aspect but we will  
focus instead  on the scaling beaviour of the ratios $P_{k}(N) / P_{\emptyset}(N) = Z_{k}(N) / Z_{\emptyset}(N)$.

It is important to stress that the exponent $\alpha_k = \alpha_{\emptyset} + m_k$ is consistent with the recent 
finding that prime knots in SAPs are weakly localized, i.e  
they  occupy on average a portion of the ring that scales as 
$N^t$, with $t\simeq 0.7$~\cite{Marcone:2005:J-Phys-A}. 
Indeed, given that each prime knot is weakly localized, SAPs with knot type $k$ should, 
in the large $N$ limit, look like as unknotted rings with
$m_k$ decorated vertices. These decorations can sit in $\sim N$ positions 
along the $N$-step SAP. The partition function of such decorated chain
thus includes a scaling law $N^{m_k}$ multiplying the partition function of unknotted chains, 
whose power-law part scales $\sim N^{\alpha}$.
Similarly, the average extension of each prime component should not contribute, in the large 
$N$ limit, to the overall extension of the knotted SAP and one would expect that SAPs with a 
fixed knot type $k$ would share, to leading order, the same metric properties of unknotted 
SAPs~\cite{Orlandini:1998:J-Phys-A}.

A more recent numerical calculation of SAPs on the cubic lattice 
has confirmed and improved the above conjectures by 
suggesting the validity of the following scaling laws~\cite{Baiesi:2010:JSM}
\begin{equation}
 Z_k(N) \simeq Z_\emptyset(N) \times \frac{N^{m_k}}{C_k},
\label{scal_ck}
\end{equation}

where $C_k$ is a coefficient growing with the knot complexity and
factorizable into the elementary contributions of each prime component. 
Roughly speaking $C_k$ can be interpreted as the elementary entropic cost to 
tie a given knot $k$ in a SAP and its dependence on $k$ can be related to 
the minimal 
number of steps necessary to build a knot type $k$ on the underlying lattice~\cite{Baiesi:2010:JSM}. 
Note that, for any two prime knots $k_1$ and $k_2$, the inverse proportionality  relation
$A_{k_1}/A_{k_2} = C_{k_2}/C_{k_1}$ is satisfied. 

One of the aims of this work is to explore more deeply the scaling relation 
(\ref{scal_ck}) by extending the numerical investigations 
in~\cite{Baiesi:2010:JSM} to SAPs embedded both on BCC and FCC 
lattices and to look at the dependence of $C_k$ on these lattices.
While the value of  $C_k$ should depend on the three-dimensional lattice considered
we expect the ratios $Z_{k_1}(N)/Z_{k_2}(N)$  to be independent on lattice 
details.
This feature has been already suggested in~\cite{Rensburg&Rechnitzer:2011:JPA} 
on the basis of a stochastic enumeration of $N$-steps SAPs performed with
the GAS algorithm~\cite{Rensburg&Rechnitzer:2009:JPA}, 
an extremely efficient method to sample SAPs of moderate lengths and with fixed knot type. 
Note that the conclusions given in~\cite{Rensburg&Rechnitzer:2011:JPA} 
rely on an extrapolation to infinity of the data obtained for SAPs within the range $N\le 500$, 
i.e. in a region of $N$s where, especially for knotted configurations, 
strong corrections to scaling are expected ~\cite{Baiesi2010:PRE}.
   
In this work we extend the range of sampled $N$ up to $100000$ by means of efficient Monte Carlo samplings.
We have been able to compute the amplitudes $C_k$ for different knot types and to confirm, 
for a wide range of  $N$, the universal character of the ratios $Z_{k_1}(N)$ / $Z_{k_2}(N)$ 
found in~\cite{Rensburg&Rechnitzer:2011:JPA} although with different asymptotic values. 

In addition, by computing the mean squared radius of gyration of each SAP, we  confirm,  
within a wide range of $N$, the validity of the scaling law (\ref{scal_radius_knot}).
Finally, as a measure of the geometrical entanglement of knotted SAPs we have considered their writhe.
Rigorous arguments  have shown that,  for SAPs on the cubic lattice, the mean absolute writhe 
$\langle |Wr| \rangle$  increases at least as 
rapidly as $\sqrt{N}$~\cite{Buks:1993:J-Phys-A} and numerical estimates 
on the same system gives $\langle |Wr| \rangle \sim N^{\eta}$ with 
$\eta = 0.5035 \pm 0.0006$~\cite{Baiesi_et_al:2009:JCP}. 
It is then natural to check whether, for SAPs with fixed knot type $k$,  the scaling law for the spread of the writhe 
and the exponent $\eta$ are independent either on the knot type $k$ or on the underlying lattice.

\section{Models and methods}

\begin{figure*}[t]
\begin{center}
\includegraphics[angle=0,width=3.6cm]{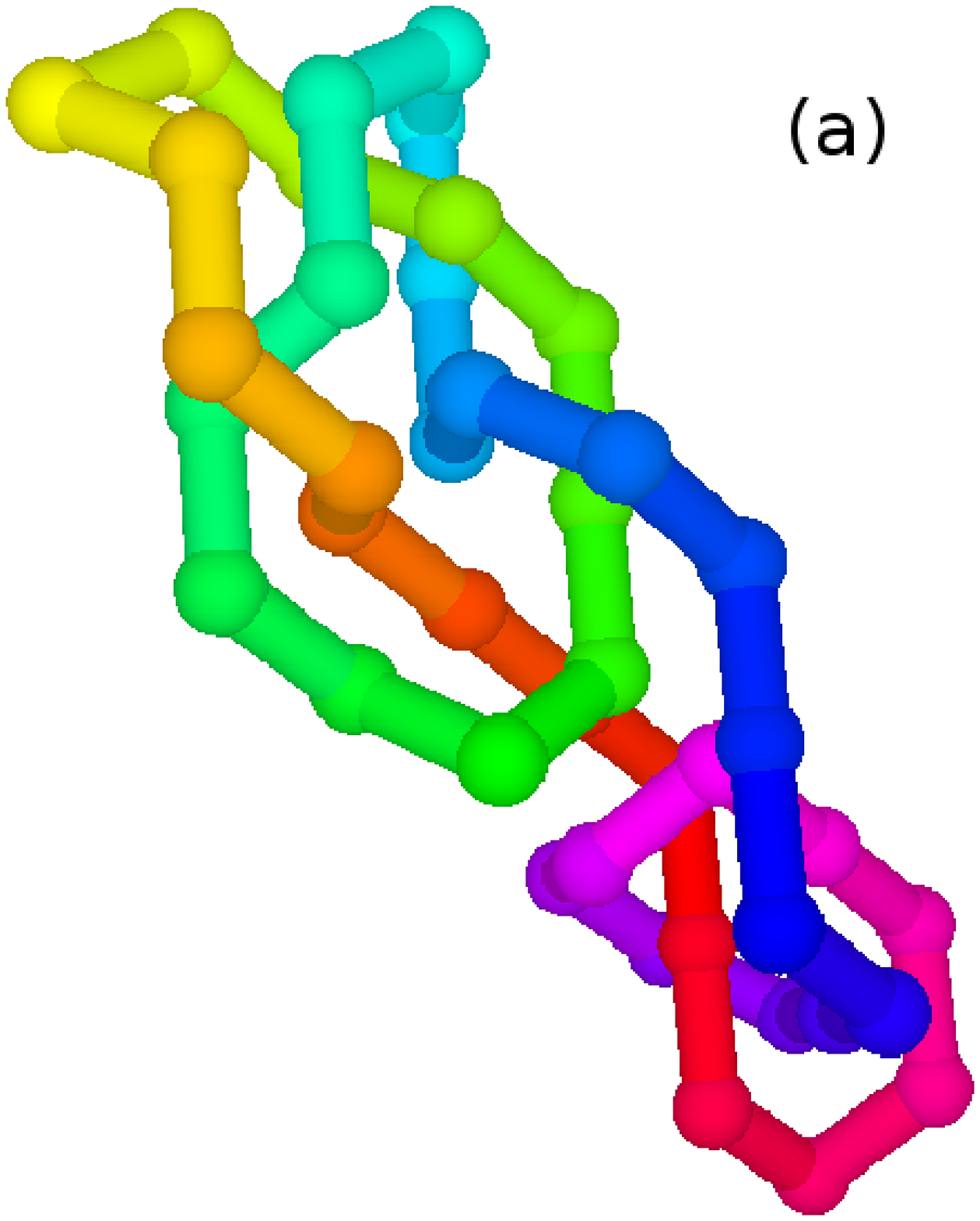}
\hskip 1mm
\includegraphics[angle=0,width=3.7cm]{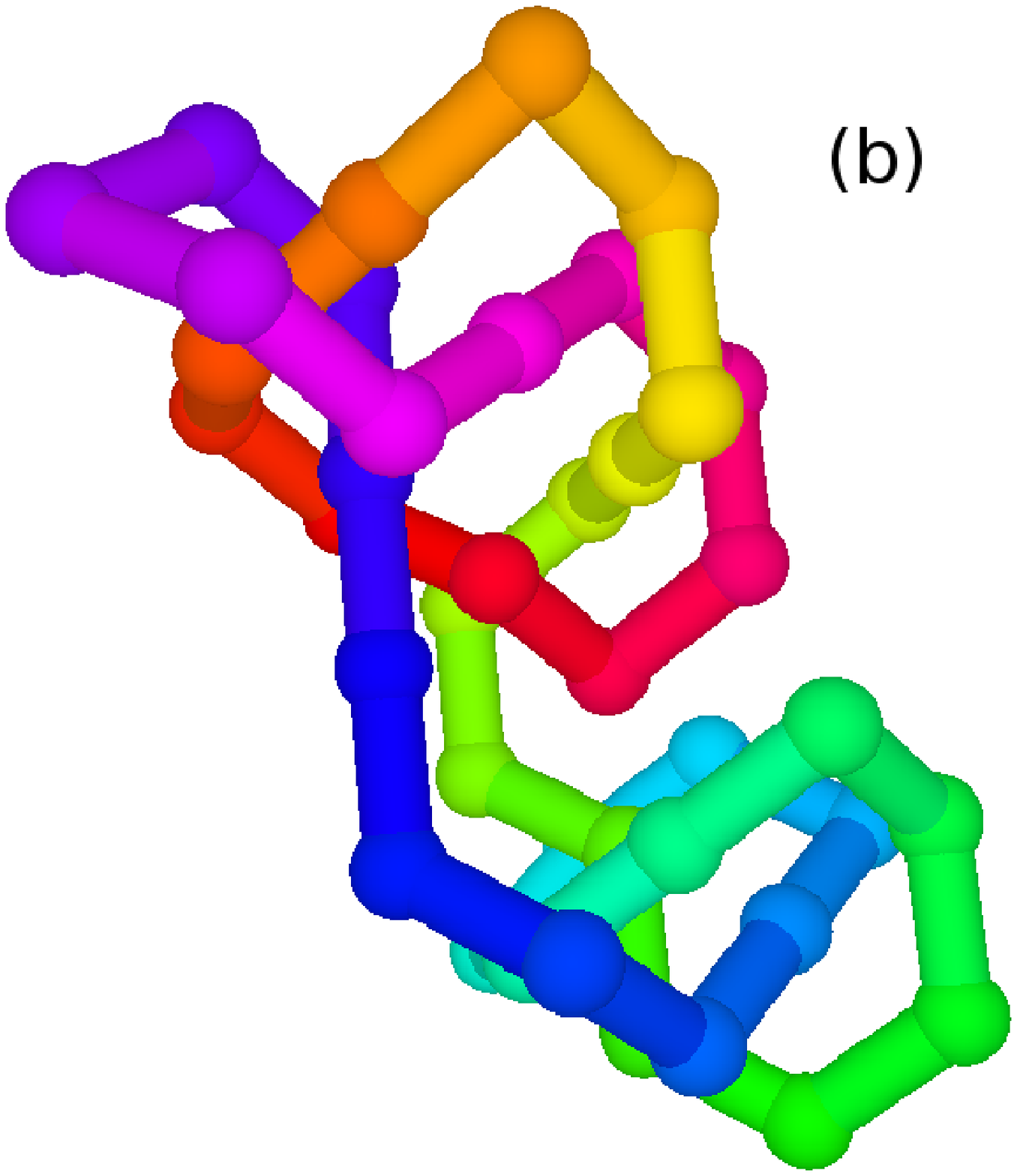}
\hskip 5mm
\includegraphics[angle=0,width=4.2cm]{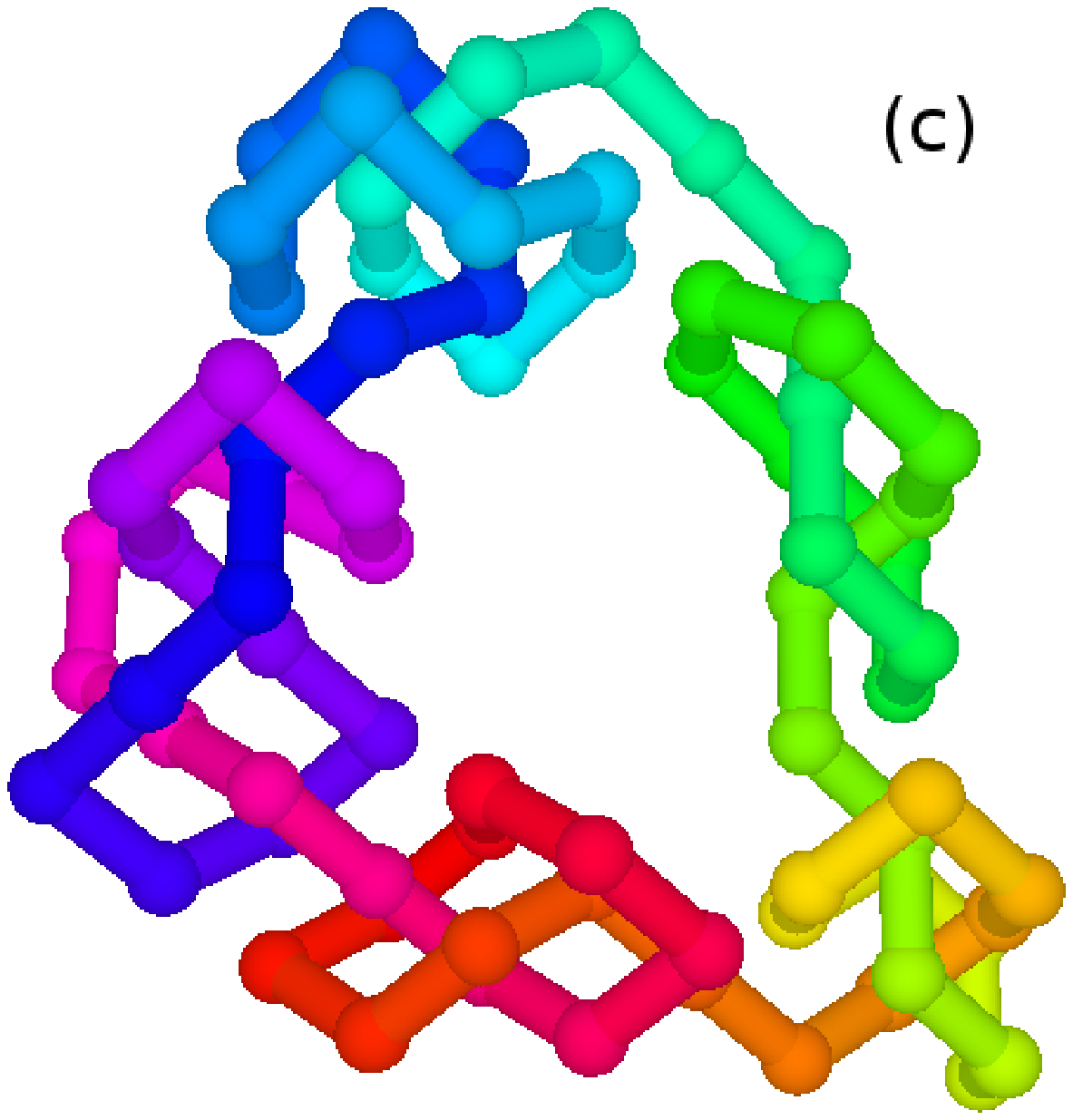}
\end{center}
\caption{(Color online) 
Examples of configurations obtained by the BFACF procedure that reduced their length from 
$N=100000$ to a few steps, with the hue that follows the monomer sequence: 
(a) $3_1$\#$5_1$ knot on the BCC lattice, 
(b) $3_1$\#$5_1$ knot on the FCC lattice, and
(c) a very tight configuration with knot type $(3_1)^4$\#$4_1$ on the BCC lattice. 
\label{fig:ex}}
\end{figure*}
 
We perform Monte Carlo sampling of SAPs on the SC, BCC and FCC lattices by using the 
two-point pivot moves, a fixed-$N$ algorithm that has been 
proven to be ergodic in the class of all SAPs and shown to be 
efficient in sampling uncorrelated configurations~\cite{Madras&Slade:1993,Madras_et_al:1990:JSP,Buks_et_al:1990:JPA}.
This algorithm allows us to reach values of $N$ up to $10^5$ for the BCC and FCC lattice.
For the SC lattice instead we use previous data~\cite{Baiesi:2010:JSM} including
also $N=1.5\times 10^5$ and  $N=2\times 10^5$.
For the longest chains the samples include $\approx 2\times 10^4$ independent configurations.
For $N=10^4$ this number raises to $\times 10^6$.
Typically $>10^5$ data are present for each $N$.

Since the pivot algorithm samples SAPs with any topology, it  requires 
a post processing procedure to characterize the knot type of each configuration. 
This is can be determined, for example,  by computing a 
topological invariant such as the Jones or the HOMFLY polynomials~\cite{Adams:1994}. 
In general the computational complexity
of these invariants increases exponentially with the geometrical entanglement of the curve and, 
for very large $N$,  this could be the most problematic part of the whole investigation. 
To overcome this problem we first smooth  each configuration by reducing its length while keeping the knot type 
fixed. This is achieved by using a nonequilibrium stochastic scheme based on local moves
(a nonequilibrium variant of the BFACF method~\cite{Berg&Foester:1981:Phys-Lett-B,Aragao:1983:Nucl-Phys-B,Janse-van-Rensburg&Whittington:1991b:J-Phys-A})
 that trims recursively all kinks in the SAP. 
In addition the algorithm performs some random local rearrangements to reduce 
the chance that a fast drop in chain length results into configurations which it is 
then very unlikely to escape from. This occurs for example  when multiple prime components are 
present in the SAP and the resulting frozen configurations are characterized by having 
these components separated by stretched portions of the SAP that become difficult to shrink. 
We find that in most situations the algorithm is able to reach a configuration close to the minimal 
$N$ allowed by its knot type on that specific lattice. 
This  configuration is then projected  along a given (approximately) irrational direction and 
 the resulting knot diagram is mapped into the corresponding 
Dowker code~\cite{Adams:1994} 
that is eventually further simplified  and factorized into the  Dowker codes of the prime knots. 
Finally,   we compare each component of the original Dowker code  against a look-up table of Dowker codes of prime knots of up to 11 crossings. In this way we have been able to distinguish 
composite knots with up to 5 prime components and with each prime component having minimal crossing number up to 11.

As an example of the efficiency of the simplification algorithm  we show  in Fig.~\ref{fig:ex}  the result of the step reduction applied to configurations with $N=100000$.
The first two configurations display a $3_1$\#$5_1$ knot (i.e.~a composite knot formed by a prime knot $3_1$ and
a prime knot $5_2$), (a) on the BCC and (b) on the FCC lattice. 
One can readily see the  $3_1$\#$5_1$ knot at these reduced lengths: 
the $3_1$ part is on the lower-right side and the $5_1$ part on the upper-left one.
The configuration in Fig.~\ref{fig:ex}(c) shows  instead a $(3_1)^4$\#$4_1$ knot (this a shorthand
notation for $3_1$\#$3_1$\#$3_1$\#$3_1$\#$4_1$).  
All these examples  confirm  the efficiency of the BFACF method in shrinking the 
chain from a very long length to a very short one, even for complex composite knots.

\section{Results}

\subsection{Frequencies of prime knots}

\begin{table}[b]
\begin{tabular}{|l|l|l|l|l|l|}
\hline 
         &\multicolumn{3}{|c|}{ $C_k / 10^5$ } & \multicolumn{2}{|c|}{ $C_k^{\rm SC}/C_k^{\rm L}$ }\\ 
         \cline{2-6} 
knot     &       SC  &          BCC     &       FCC     &   L=BCC & L=FCC   \\
\hline
$3_1$    &       2.28(2) &      1.57(1) &       1.64(2) &       1.45(2) &   1.39(3)\\
$4_1$    &       50(1)  &       35.5(5) &       37.5(5) &       1.41(5) &   1.33(5)\\
$5_1$    &       510(30) &      378(18) &       380(10) &       1.35(14) &  1.34(12)\\
$5_2$    &       334(15) &      233(7)  &       257(10) &       1.43(11) &  1.30(11)\\
$6_1$    &       5700(1200) &   3000(500) &     3100(400) &     1.9(7)   &  1.84(63)\\
$6_2$    &       6800(1900) &   3450(600) &     3500(400) &     2.0(9)   &  1.94(77)\\
$6_3$    &       6900(1300) &   5900(700) &     6000(950) &    1.17(36) &  1.15(41)\\
\hline
\end{tabular}
\caption{Knot costs $C_k$ for the SC, BCC, and FCC lattices, determined from averages of the
data $N Z_\UN/Z_k$ in the range $5000\le N<100000$ (specific points clearly nonasymptotic for the knots
with $6$ crossings have also been excluded from the averages). Errors refer to one standard deviation. 
The ratios $Z_k^{\textrm{BCC}} / Z_k^{\textrm{SC}}$ and 
$Z_k^{\textrm{FCC}} / Z_k^{\textrm{SC}}$ in the last two columns seem to be independent on the knot $k$.
}.
\label{tab1}
\end{table}

\begin{table}[b]
\begin{tabular}{|l|l|l|l|l|l|}
\hline 
         & &\multicolumn{4}{|c|}{ $Z_{k_1}/Z_{k_2}= C_{k_2}/C_{k_1}$ }\\ 
         \cline{3-6} 
$k_1$ & $k_2$    &       SC  &          BCC     &       FCC      & mean\\
\hline
$3_1$ & $4_1$    &       21.9(7)  &      22.6(5) &     22.9(6)   & 22.5(4)   \\
$3_1$ & $5_1$    &       224(15)  &      241(13) &     231(9)    & 232(9)  \\
$3_1$ & $5_2$    &       146(8)   &      149(6)  &     157(8)    & 150(5)   \\
$3_1$ & $6_1$    &       2500(550) &   1900(350) &     1900(300) & 2000(300)   \\
$3_1$ & $6_2$    &       3000(900) &   2200(400) &     2100(300) & 2200(400)   \\
$3_1$ & $6_3$    &       3000(600) &   3750(500) &     3650(650) & 3500(400)   \\
\hline
\end{tabular}
\caption{Ratios of partition function of $3_1$ and of other simple prime knots $k_2$ (i.e. $C_{k_2}/C_{3_1}$), 
for the three lattices, showing independence on the lattice type.}
\label{tab2}
\end{table}

\begin{figure}[t]
\begin{center}
\includegraphics[angle=0,width=8.2cm]{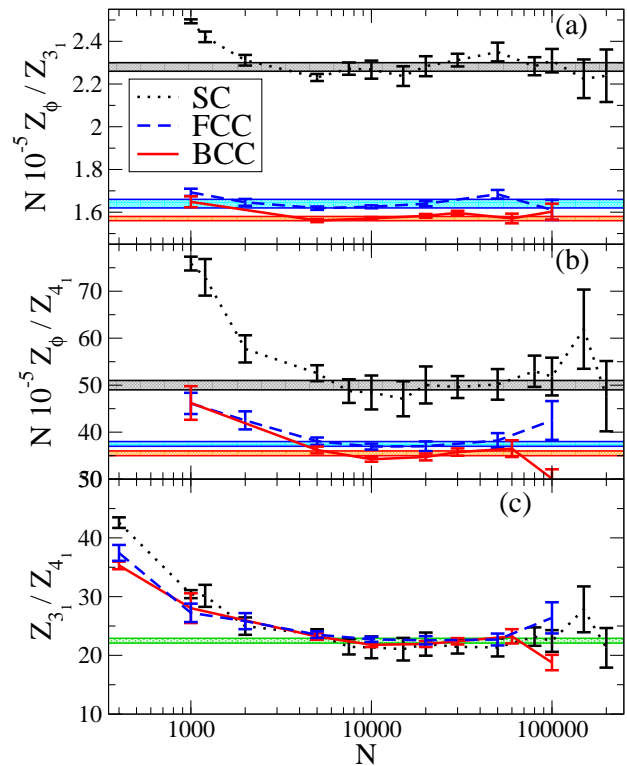}
\end{center}
\caption{(Color online)
(a) Ratio of the probability of unknot configurations and of $3_1$ knots, times $N\times 10^{-5}$.
They converge to a lattice-dependent constant ($C_{3_1}\times 10^{-5}$) 
for large $N$. Bands of different colors represent the estimated asymptotic values with $\pm$ error,
see Table~\ref{tab1}.
(b) The same for knot $4_1$. 
(c) Ratio of probability of $3_1$ knot and of $4_1$ knot, converging to
a constant for large $N$ ($C_{4_1}/C_{3_1}$) 
that does not depend on the lattice kind. The horizontal lines mark the interval
of the estimated asymptotic value, see Table~\ref{tab2}. This panel includes also data at $N=400$.
Note that, while the small $N$ deviations from the asymptotic values are due to corrections to scaling,
the fluctuations at larger $N$'s are originated by limited statistics. 
\label{fig:41}}
\end{figure}

\begin{figure}[t]
\begin{center}
\includegraphics[angle=0,width=8.2cm]{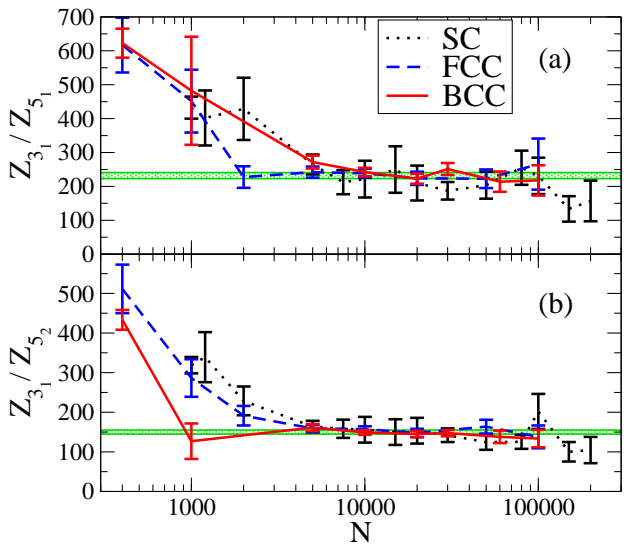}
\end{center}
\caption{(Color online) 
Ratio of probability of $3_1$ knot and of (a) $5_1$ knot, (b) $5_2$ knot.
The horizontal lines represent the estimated asymptotic ratio 
(Table~\ref{tab2}, last column) $\pm$ error.
\label{fig:5152}}
\end{figure}

The BCC and the FCC lattices have coordination number $8$ and $12$, respectively.  
It could be expected that more neighbors implies more flexibility and thus a more pronounced tendency to form 
knots. A manifestation of this effect would be that the  knot costs $C_k$ were somewhat inversely
proportional to the coordination number.
It turns out that indeed knots are less frequent in SAPs embedded on the SC lattice and hence
$C_k^{\textrm{SC}}$ is larger than its counterparts for FCC and BCC lattices. 
However, for the simplest knots we observe surprisingly that $C_k^{\textrm{BCC}} < C_k^{\textrm{FCC}}$.
According to (\ref{scal_ck}) this is readily seen by determining $C_k$ from $N Z_\UN / Z_k$
for a simple prime knot.
For instance, by focusing on  $C_{3_1}$  in  [Fig.~\ref{fig:41}(a)] and $C_{4_1}$ 
in  [Fig.~\ref{fig:41}(b)] it is easy to  see that these quantities are smaller  for SAPs on
the BCC lattice than on the FCC.
If we consider instead the ratio $Z_{3_1}/Z_{4_1}$ [Fig.~\ref{fig:41}(c)] we note that, 
as $N$ increases,  it  approaches an asymptotic value that is  independent  on  the 
lattice. This is true also for frequency ratios between other prime knots, as shown 
in Fig.~\ref{fig:5152} and in Table~\ref{tab2}.   
These results support the previously conjectures hypothesis 
that the frequency ratios between prime knots do not depend on the lattice in which they are embedded~\cite{Rensburg&Rechnitzer:2011:JPA}. Our estimates in Table~\ref{tab1} and in Table~\ref{tab2}  differ however from the ones reported in ~\cite{Rensburg&Rechnitzer:2011:JPA}. A possible explanation is that the estimates in~\cite{Rensburg&Rechnitzer:2011:JPA} are  affected by  systematic errors due to the small values of $N$ considered 
whose range fall  in a region where the corrections to scaling are too strong (see the data for $N=400$ in Fig.~\ref{fig:41}(c)) to be neglected.

Another way of presenting these results is by noticing  that the frequencies of prime knots, 
whose inverse is proportional to the costs $C_k$, have fairly fixed ratios for different lattices:
according to columns 5 and 6 of Table~\ref{tab1}, we see indeed that for each knot
on the SC there are $1.44(3)$ knots on the BCC, and $1.37(3)$ knots on the FCC 
(these values correspond to weighted averages of the columns assuming a constant 
knot-independent ratio for the two  lattices).

\subsection{Frequencies of composite knots}
The case of composite knots formed by $m_k=m$ topologically equivalent prime knots $k$
(denoted $(k)^m$ hereafter), was recently discussed in~\cite{Baiesi:2010:JSM} where 
the entropic cost  $C_{(k)^m}$ was determined in a rather simple way in terms of the
corresponding costs necessary to tie  its prime components. 
Since each component is  (weakly) localized,  in the large $N$ limit
we can think of the chain with knot   $(k)^m$ as an  unknotted ring decorated by 
topologically identical objects (prime knots), each placed in one of  $\sim N$ available locations.
The entropic cost of each decoration is  $C_k$, and being all independent  one would 
expect  that the full cost $C_{(k)^m}$ is simply given by  the product of the $m$ elementary costs  $C_k$. 
However, in this picture, being the prime knots topologically identical,  they can be interchanged along the chain keeping the configuration undistinguishable. This property gives rise to  a   combinatorial factor $m!$ that needs to be removed from the counting in order to define the 
relative entropy of the  decorated chain with respect to the unknotted one.
This gives
\begin{equation}
C_{(k)^m} = m!\, (C_k)^m.
\label{cristal}
\end{equation}
For more general composite knots that include  groups of different prime knots, 
the total cost is the  product of the cost of each group given by  (\ref{cristal}).

\begin{figure}[!tpb]
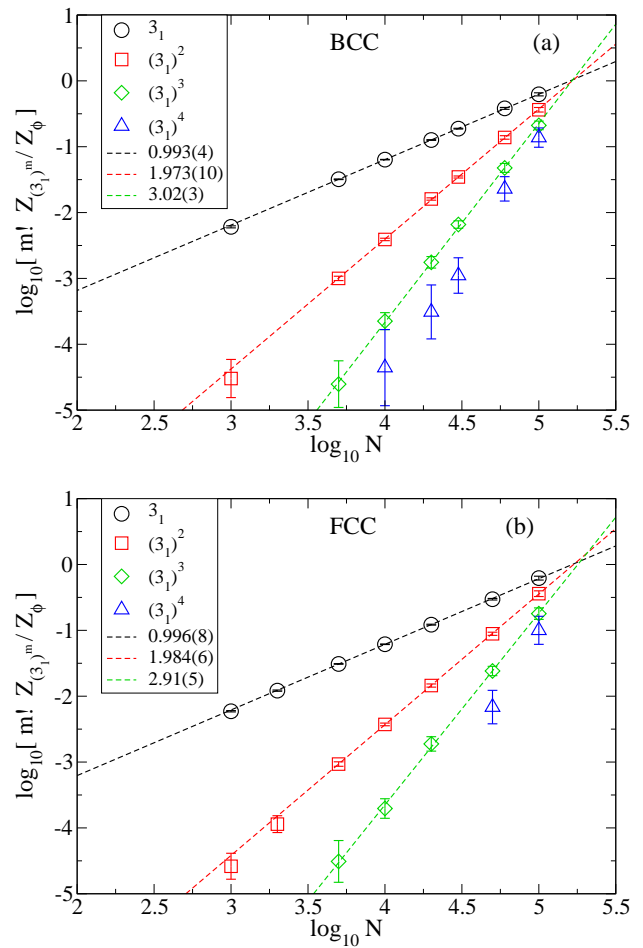

\begin{center}
\includegraphics[angle=0,width=8.2cm]{fig_bccfcc_4a.eps}
\vskip 3mm
\includegraphics[angle=0,width=8.2cm]{fig_bccfcc_4b.eps}
\end{center}
\caption{(Color online) 
Log-log plot of ratios of partition functions $Z_{(3_1)^m}/Z_\UN$ vs.~$N$,
for multiple trefoil knots, in (a) the BCC lattice and (b) the FCC lattice.
The straight lines are fits compatible with power laws $N^m$ (exponents are in the legend). 
For each lattice, the crossing of these power-laws is at a single point of coordinates $(C_{3_1}, 1)$.
\label{fig:cristal}}
\end{figure}

For trefoil knots on the SC lattice is was shown that (\ref{cristal}) 
was fulfilled by an excellent degree of precision~\cite{Baiesi:2010:JSM}. 
In Fig.~\ref{fig:cristal} we can see that also on FCC and BCC
lattices the data follow the relation $Z_{(3_1)^m}/Z_\UN = N^m C_{(k)^m}$ 
with $C_{(k)^m}$ given by (\ref{cristal}).

The importance of the factorial term $m!$ can be understood as follows:
Since the trefoil is the prime knot with the lower cost $C_k$, 
one would first guess that SAPs with the  composite knot $(3_1)^{m}$ 
are the most abundant ones for $N>C_{3_1}$.  However, when the cost ratio $C_{3_1}/C_{4_1}$ 
becomes smaller than $m$, because of the factorial factor $m$!,   configurations with knot  $(3_1)^{m-1}$\#$4_1$ are more frequent than those hosting the  knot $(3_1)^{m}$.
Since $C_{3_1}/C_{4_1} \approx 22$ this means that the combinatorial entropy loss 
should start to become relevant for a number of prime components  $m\gtrsim 22$, 
which corresponds to $N\approx (C_{3_1})^{22}$, i.e.~chain lengths
(for example, $N\approx 10^{114}$ on the BCC lattice) that are impossible to test numerically.
Hence, we can safely expect trefoils to dominate the knot statistics  for realistic polymer rings in good solvent regime.

\subsection{Metric properties}

\begin{figure}[!tpb]
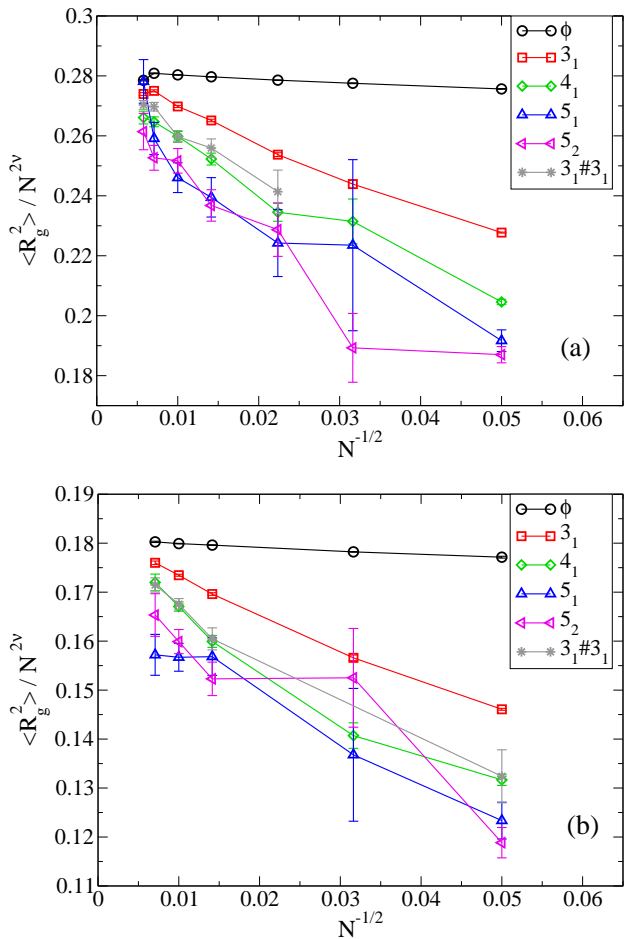

\begin{center}
\vskip 2mm
\includegraphics[angle=0,width=8.2cm]{fig_bccfcc_5a.eps}
\vskip 3mm
\includegraphics[angle=0,width=8.2cm]{fig_bccfcc_5b.eps}
\end{center}
\caption{(Color online) 
Mean radius of gyration $\mean{R_g^2}$ divided by $N^{2\nu}$, with $\nu=0.587597$, 
i.e.~corrections to scaling of $\mean{R_g^2}$ as a function of $N^{-1/2}$: (a) BCC lattice and (b) FCC lattice. 
Note that error bars and fluctuations for knots $5_1$ and $5_2$ at $N=1000$ are larger than other ones because of worse statistics.
\label{fig:Rg}}
\end{figure}

By collecting sampled configurations with the same knot type we can compute, for example, 
the mean squared radius of gyration $\mean{R_g^2(N)}_k$ of SAPs with fixed topology 
and see if its scaling behavior depends either on $k$ or on the lattice in which the SAPs are embedded.
According to  (\ref{scal_radius_knot}) we consider the 
ratio $\mean{R_g^2(N)}_k / N^{2\nu}\sim B_k (1+b_k/N^\Delta)$ and assume $\Delta=1/2$ that is close
to all present estimates~\cite{Clisby:2010:Phys-Rev-Lett}.
In Fig.~\ref{fig:Rg} we report these ratios  as a function of $N^{-1/2}$ for 
$\nu=0.587597$~\cite{Clisby:2010:Phys-Rev-Lett} and for different knot types: 
one can readily  see  that these ratios converge (as $N \to \infty$)
to a common value,  independent on the knot type. This supports the expectation
that $B_k$ is the same for all knots~\cite{Orlandini:1998:J-Phys-A} and confirms indirectly   the 
weak localization  property of the knotted portion of the chain~\cite{Marcone:2005:J-Phys-A}.
The value of $B_k$ does however depend on the chosen lattice and is larger on the BCC 
[$\approx 0.28$, see Fig.~\ref{fig:Rg}(a)] than on the FCC [$\approx 0.18$, see Fig.~\ref{fig:Rg}(b)]. 
This is somehow expected since $B_k$ is an amplitude.
Note that, while $B_k$ does not depend on topology, this is not true 
for the amplitude of the correction to scaling $b_k$ whose value increases  with knot complexity.

\subsection{Writhe}

\begin{figure}[!tpb]
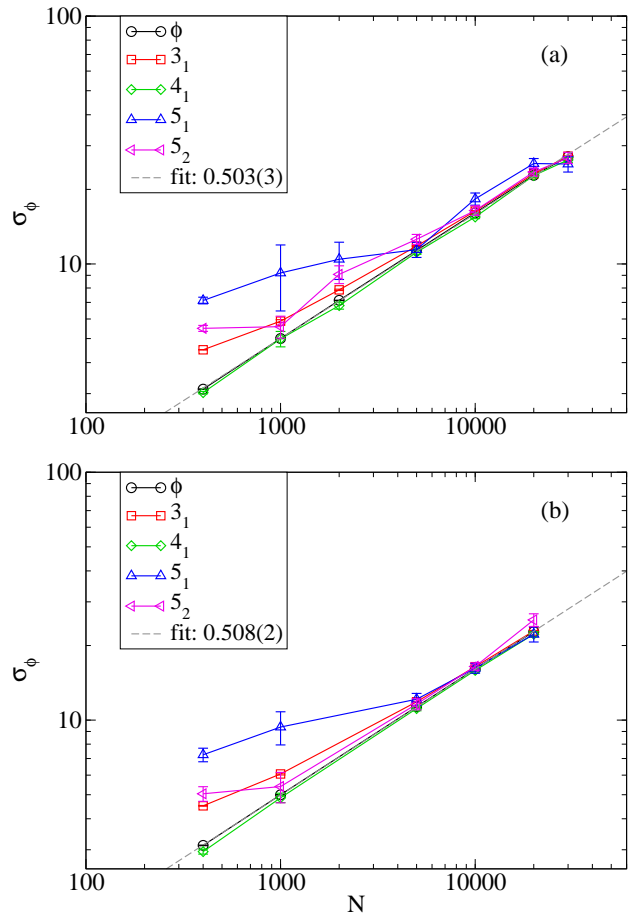

\begin{center}
\vskip 2mm
\includegraphics[angle=0,width=8.2cm]{fig_bccfcc_6a.eps}
\vskip 3mm
\includegraphics[angle=0,width=8.2cm]{fig_bccfcc_6b.eps}
\end{center}
\caption{(Color online) 
Log-log plot of the standard deviation of the writhe distribution as a function of $N$: 
(a) BCC lattice and (b) FCC lattice. 
\label{fig:sigma_Wr}}
\end{figure}

The writhe of a curve is a quantity that describe its geometrical self-entanglement. 
A commonly used algorithm to compute the writhe of a curve goes as follows: First, one projects the
curve onto an arbitrary plane. In general the projection will have crossings that most of the time
will be transverse, so that, after having established an orientation of the curve, a sign $+1$ or
$-1$ (determined by a right hand rule)
can be assigned to each crossing. The sum of these signs  gives the signed crossing number
in this projection. The writhe of the curve is obtained by averaging these signed crossing
numbers over all possible projections. From this definition it is clear that the main difficulty in
computing the writhe of a configuration would be the averaging procedure over all projections.
Fortunately, for polygons on SC, FCC and BCC this procedure is enormously simplified by theorems
\cite{Lacher_Sumners:1991,Garcia_et_al:1999:JPA,Laing_Sumners:2006:JPA} 
which reduce the computation of writhe to the average of 
linking numbers of the given curve with a finite set of selected push-offs of the curve itself. 
In our calculation of the writhe we made extensive use of these results.

Clearly the writhe of SAPs is, on average, zero and, provided we do not distinguish mirror images in chiral knots, this is still true for SAPs with fixed knot type. 
The simplest non trivial observable for the writhe  distribution of SAPs with knot type $k$ is then its standard deviation $\sigma_k$.
In Fig.~\ref{fig:sigma_Wr} we report the log-log plot of several $\sigma_k$ as a function of $N$. 
Different symbols refer to different knot types and it is readily seen that, for sufficiently 
long $N$ all data converge to a common curve that is essentially the one for unknotted SAPs.
We can then argue that, similarly to SAPs in the SC lattice~\cite{Baiesi_et_al:2009:JCP},  and for all the knots considered here  $\sigma_k \sim \sigma_\UN = D_\UN\,N^{\eta_\UN}$, both for FCC and BCC lattices.

A simple linear fit of the log-log data gives estimates $\eta_{\UN} = 0.503(3)$, $D_{\UN} = 0.155(5)$  for BCC and $\eta_{\UN} = 0.508(2)$, $D_{\UN} = 0.1495(14)$ for the FCC. 
These estimates are quite consistent with the estimate $\eta_{\UN}=0.506(1)$ found for the SC lattice~\cite{Baiesi_et_al:2009:JCP}.  This is not true, however, for the amplitude $D_{\UN}$ whose estimate in the SC ($D_{\UN} = 0.1369(7)$) differs from the ones shown above for the other two lattices.

Similar results are found for the absolute value of the writhe (exponent 
$\eta_\UN=0.504(4)$ on the BCC and $\eta_\UN=0.508(2)$ on the FCC), which is expected, given the regular shape
of the writhe distribution~\cite{Baiesi_et_al:2009:JCP}.

Two features are worth noticing: firstly the 
estimate of $\eta_\UN$ is very close to the lower bound $1/2$ proved rigorously for SAPs with free topology 
in the SC lattice~\cite{Buks:1993:J-Phys-A}.
Secondly the deviation from the scaling law experienced by SAPs with fixed knot type
is less pronounced for the figure-eight and the unknot (achiral) then for all the others knots that are chiral.
This is related to the nonzero mean writhe of each of the two images of a chiral knot, 
see~\cite{Baiesi_et_al:2009:JCP}.

\section{Discussion}
Since renormalization group arguments are not applicable to the statistical ensembles of 
rings with fixed knot type, we have resorted to simulations to investigate the  universal features of the scaling laws for the entropic, metric and geometrical properties of knotted SAPs. 
We have confirmed that frequencies of knots in SAPs depend on the lattice in which the configurations are embedded,  On the other hand  the ratios between knot frequencies are  lattice-independent numbers and depend only on the knot types involved.  Surprisingly, knotted configurations are more frequent on the BCC than on the FCC lattice. This finding is unexpected
because the FCC lattice has a higher coordination number than the BCC one, and because the knots with the shortest number of steps are found in the FCC lattice~\cite{Rensburg&Rechnitzer:2011:JPA}. 

We have also supported the conjecture that composite knots in SAPs appear  with frequencies that, 
in the large $N$ limit, can be
inferred from those of the prime knots in the knot decomposition.  This property  was conjectured  
for polygons on the cubic lattice~\cite{Baiesi:2010:JSM}  and  here is confirmed also for SAPS on the  BCC and FCC lattices.
The composite \emph{knot cost} includes a factorial $m!$ for every group of $m$ identical prime knots, 
which multiplies the $m$-th power of their cost. The factorial term is suitable to properly take into account the 
combinatorial entropic loss for identical prime components in the picture where, in the large $N$ limit,  each prime component  
behaves as a decorating  point along the unknotted ring.  
As a byproduct one can predict that,  for  chain lengths $N>10^{114}$,
knots  with the highest frequency include not only trefoil knots but also other  prime knots.

Concerning the metric properties of SAPs with fixed topology,  our results for all the three lattices considered confirm that  not only does the mean squared radius of gyration share the usual power law $N^{2\nu}$  known for the set of all SAPs  but also  the  amplitude $B_k$ does not depend on the knot type $k$~\cite{Orlandini:1998:J-Phys-A},
 see (\ref{scal_radius_knot}).  The dependence on the knot type is  however present at the level of the  corrections to scaling, which become more pronounced as the knot complexity of the SAPs increases. 

Finally we have  shown that the   large $N$ scaling behavior of the variance of the writhe (or equivalently  its absolute value) for SAPs with  knot type $k$  is independent both on $k$ and lattice type and is very close  to $\sqrt{N}$,   which is the lower bound  proved 
by rigorous arguments  for the class of all SAPs on the SC lattice~\cite{Buks:1993:J-Phys-A}.

By universality it is reasonable to expect that most of the  features shown here for  SAPs on  lattices can be valid also for other  models  of polymer rings in which the excluded volume interaction is taken into account.  On the other hand preliminary studies on the knot probability  for off-lattice rings have either neglected completely the excluded volume interaction~\cite{Dobay:2003:PNAS} or looked at the 
knot probability amplitudes for  few and rather small values of $N$~\cite{Deguchi&Tsurusaki:1997:PRE} . 
In the near future it would be then interesting to extend the analysis reported here to a larger set of  polymer models.


%

\end{document}